\documentclass[aps,prd,preprint,superscriptaddress,nofootinbib]{revtex4-2}

\usepackage{amsmath}
\usepackage{amssymb}
\usepackage{hyperref}
\usepackage{graphicx}
\usepackage{dcolumn}
\usepackage{bm}
\usepackage{mathrsfs}

\DeclareSymbolFont{usualmathcal}{OMS}{cmsy}{m}{n}
\DeclareSymbolFontAlphabet{\mathcal}{usualmathcal}

\newcommand{\mA}{m_A}
\newcommand{\mB}{m_B}

\usepackage{xcolor}
\usepackage[normalem]{ulem}
\begin{document}
\global\count\footins=1000 

\preprint{MIT-CTP/6097}

\title{A purely mechanical system realizing a Coulomb-like interaction
}

\author{June-Haak Ee}
 \email[]{jhee@mit.edu}
 \affiliation{Center for Theoretical Physics -- a Leinweber Institute, Massachusetts Institute of Technology, Cambridge, MA 02139, USA}
\author{U-Rae Kim}%
 \email[]{kim87@kma.ac.kr}
 \affiliation{Department of Physics, Korea Military Academy, Seoul 01805, Korea}%
\author{Jungil Lee}
 \email[]{jungil@korea.ac.kr}
 \affiliation{Department of Physics, Korea University, Seoul 02841, Korea}%
 \affiliation{Quark Valley, Gagok-Ro 163, Namyangju 12032, Korea}%
\collaboration{\textsl{KPOP}$\mathscr{E}$ Collaboration}

\date{\today}

\begin{abstract}
We solve in closed form a one-dimensional relativistic system: two masses
interacting only through elastic collisions with a massless mediator
bouncing between them. Momenta, times, and positions are hyperbolic
functions of the collision index. The mediator energy, interpreted as the
pair's effective potential, obeys an exact discrete Coulomb law,
$V\propto 1/r$, with a Lorentz-invariant action as coupling. A massive
Newtonian mediator instead transmits a $1/r^{3}$ force; one
adiabatic invariant traces both laws to the mediator's dispersion
relation. Continued to negative mediator energy, the closed
forms turn trigonometric, binding a one-dimensional mechanical analog of
the Coulomb atom.
\end{abstract}

\maketitle


\section{Introduction}
\label{sec:intro}

One-dimensional elastic collisions are among the most instructive exactly
solvable problems of classical mechanics: energy and momentum conservation
alone determine the final state of every two-body collision, and chains of
such collisions generate surprisingly rich dynamics~\cite{Ee2012,Ee2015}.
In a previous work~\cite{Ee2012}, we studied a block of mass $M$ that
repeatedly collides with a light ball of mass $m$ sandwiched between the block
and a rigid wall. The complete trajectory of the block was obtained
analytically, and near the turning point the ball was shown to transmit an
effective force proportional to $1/r^{3}$, where $r$ is the block--wall
distance --- the same power law as the force between a point charge and an
aligned electric dipole. In a sequel~\cite{Ee2015}, we generalized the system to two identical
blocks exchanging momenta through a ball and found the \emph{magic mass
ratios} at which the energy and momentum of the incident block are completely
transferred to the target, even though the transfer proceeds through
multiple intermediate collisions.

The block-and-wall system has also attracted wide attention outside the research literature.
For the mass ratio $m/M=100^{-n}$, the total number of collisions counts the digits of
$\pi$ --- a phenomenon discovered by Galperin~\cite{Galperin2003} and
analyzed by Redner~\cite{Redner2004}, which reached an audience of millions
through the videos of 3Blue1Brown~\cite{3b1b_1,3b1b_2}. (For the dynamics of
a heavy particle among many light ones see Sinai~\cite{Sinai}; for the
history of the counting, Ref.~\cite{Aretxabaleta2020}.) The collision count is contained
in the analytic solution of Ref.~\cite{Ee2012} as a one-line corollary:
the total number of block--ball collisions is
$N=\lceil \pi/\theta-\tfrac12\rceil$ with $\tan(\theta/2)=\sqrt{m/M}$.
The ball--wall reflections alternate with these, so their number is $N$ or
$N-1$, and the grand total is
$\lfloor\pi/\arctan\sqrt{m/M}\rfloor\simeq\pi\times10^{n}$, whose leading
digits are those of $\pi$~\cite{Galperin2003} (up to the boundary case of an integer $\pi/\arctan\sqrt{m/M}$, as for $m=M$, where the count is one less).
The quantum version of the system has been connected to Grover's search
algorithm by Brown~\cite{Brown:2019jvs} and to scattering phase shifts by Cai
and Zhang~\cite{CaiZhang2023}, and the classical $\pi$-counting continues to
attract attention~\cite{Aretxabaleta2020}.

\begin{figure}[t]
\centering
\includegraphics[width=0.8\linewidth]{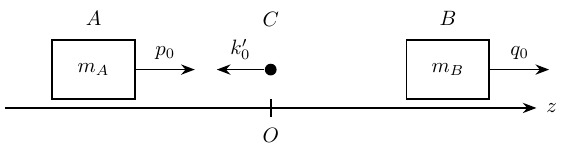}
\caption{\label{fig:model}%
The model system. Two massive particles $A$ ($\mA$) and $B$ ($\mB$) interact
only through elastic collisions with a massless particle $C$ that bounces
between them. Initially $C$ moves to the left with momentum $-|k_0'|$ toward
$A$. The figure shows the configuration right before the first $A$--$C$
collision.}
\end{figure}

In this paper we ask the natural next question in this series: \emph{what
does special relativity do to the mediated force?} Relativity admits a
qualitatively new mediator --- a \emph{massless} particle --- which has no
Newtonian counterpart. We therefore consider two massive particles $A$ and
$B$ on a line, interacting only through elastic collisions with a massless
particle $C$ that bounces between them
(Fig.~\ref{fig:model}). All three participants are treated fully
relativistically, and every collision conserves total energy and momentum.

The system turns out to be exactly solvable, and the answer to the question is
striking. The energy--momentum recurrence relations linearize in light-cone
coordinates, and all momenta after the $n$th collision are hyperbolic
functions of $n$ [Eqs.~(\ref{eq:pnm})--(\ref{eq:knabs})], mirroring the
trigonometric functions of $n$ that solve the Newtonian problem with its
massive mediator~\cite{Ee2012} --- rotation has become boost. 
Summing the flight segments yields the
collision times and positions in closed form. Interpreting the mediator
energy as the potential energy of the pair --- the same bookkeeping as in
Ref.~\cite{Ee2012} --- we find an \emph{exact discrete Coulomb law}
[Eq.~(\ref{eq:exactJ})]: with the separation sampled by the geometric mean
of adjacent flight lengths ($\tilde r$), the potential is
\begin{equation}
V(\tilde r)\;=\;\frac{J\,c}{\tilde r}
\label{eq:coulomb-intro}
\end{equation}
\emph{collision by collision}, with no continuum limit taken ($c$ is the
speed of light): the massless mediator transmits the inverse-square
repulsion.
Moreover, the coupling constant $J$ is a Lorentz-invariant quantity with
the dimension of \emph{action}, fixed by the initial conditions --- the
same structure carried by the couplings of the two long-range forces of
nature, $e_Ae_B\,e^{2}/(4\pi\epsilon_{0})$ of electrostatics and
$G\mA\mB$ of Newtonian gravity, both of dimension action$\,\times\,c$. A
single adiabatic invariant of the bouncing mediator ---
$J=\tfrac12\oint p\,dz=pr$, with $p$ its momentum, $r$ the separation of
the pair, and the loop taken over one round trip --- unifies this
result with the $1/r^{3}$ force of the Newtonian problem: 
the potential of the mediated interaction is the mediator's dispersion relation
$E_C(p)$ evaluated at $p=J/r$ (Sec.~\ref{sec:unified}). A massless mediator,
$E_C=pc$, universally yields $V\propto1/r$; a Newtonian mediator,
$E_C=p^{2}/2m$, yields $V\propto1/r^{2}$. The mechanical system thus realizes,
in an exactly solvable setting, the kinematic core of the particle-physics
folklore that long-range inverse-square forces are transmitted by massless
quanta.

This paper is organized as follows. Section~\ref{sec:model} defines the model
and the kinematic variables. In Sec.~\ref{sec:solution} we solve the
recurrence relations for the momenta in closed form and exhibit the worldline
picture of the scattering. Section~\ref{sec:times} summarizes the collision
times and positions. In Sec.~\ref{sec:potential} we derive the effective
potential and the Lorentz-invariant coupling. Section~\ref{sec:attractive}
develops the formal negative-energy branch --- the relativistic ``toy
graviton''~\cite{Artigue:2018jwu,ArtigueRel2024} --- for which the same
closed forms turn trigonometric and describe a bound pair.
Section~\ref{sec:unified}
presents the unified interpretation of the mediated force laws and discusses
the limitations of the electromagnetic analogy, and we conclude in
Sec.~\ref{sec:summary}. Derivations are collected in the appendixes.

\section{The model}
\label{sec:model}

Two massive particles, $A$ with mass $\mA$ and $B$ with mass $\mB$, and a
massless particle $C$ are aligned on the $z$ axis with $z_A<z_C<z_B$ at all
times. The only interaction is the elastic contact collision of $C$ with $A$
or with $B$; between collisions all three propagate freely. Because $C$ moves
at the speed of light it overtakes both massive particles, so the collisions
strictly alternate, $A,B,A,B,\dots$, and $A$ and $B$ never touch each
other. Every collision conserves the total energy and momentum, and the rest
masses are unchanged --- this is what we mean by \emph{elastic}. 

We denote by $A_n$ [$B_n$] the $n$th collision between $A$ and $C$
[$B$ and $C$]; its four-position is $Z_n=(ct_n,z_n)$
[$Z_n'=(ct_n',z_n')$]. 
Throughout we use the metric signature $(+,-)$ and write four-momenta as
$P=(P^0,P^3)=(E/c,\,p)$, so that $P\cdot Z$ has the dimension of action.
The four-momenta right after $A_n$ are
\begin{subequations}
\label{eq:fourmomenta}
\begin{align}
P_n&=\Big(\sqrt{p_n^2+(\mA c)^2},\;p_n\Big),
&
K_n&=\eta\big(|k_n|,\;+|k_n|\big),
\\
\intertext{and right after $B_n$,}
Q_n&=\Big(\sqrt{q_n^2+(\mB c)^2},\;q_n\Big),
&
K_n'&=\eta\big(|k_n'|,\;-|k_n'|\big),
\end{align}
\end{subequations}
where $P_n$ and $Q_n$ are the four-momenta of $A$ and of $B$, with spatial
momenta $p_n$ and $q_n$,
and $K_n$ [$K_n'$] is 
the right-moving [left-moving] four-momentum of $C$ after bouncing off $A$
[$B$]. The initial state is specified by $P_0$, $Q_0$,
the left-moving $K_0'$, and one length --- the first right-moving flight $z_1'-z_1$, which sets
the geometric scale and, with it, the coupling $J$ of Sec.~\ref{sec:potential}; the first collision is $A_1$. 
The sign $\eta=\pm1$ in Eq.~(\ref{eq:fourmomenta}) selects the branch. 
For $\eta=+1$ the mediator carries positive energy, the momentum transfer of
every collision pushes $A$ and $B$ apart, and the mediated interaction is
repulsive; this is the only branch realizable by elastic collisions of
physical particles, and Secs.~\ref{sec:solution}--\ref{sec:potential} are
devoted to it. The formal branch $\eta=-1$ --- a mediator carrying
negative energy, the relativistic analog of the ``toy gravitons''
introduced by Artigue~\cite{Artigue:2018jwu,ArtigueRel2024} --- transmits
attraction: the flip applies to the whole four-vector, so a segment whose
worldline runs to the right carries $k^{3}=-|k|<0$ as well as $E=-|k|c$,
its direction of propagation unchanged ($v=k^{3}c^{2}/E=+c$), and every
push becomes a pull. The same closed forms continue analytically to that
branch and describe attraction and, within a window of invariant mass, a
bound pair, as we show in Sec.~\ref{sec:attractive}.

It is convenient to use light-cone components,
$P^{\pm}\equiv(P^0\pm P^3)/\sqrt2$, in which the inner product reads
$P\cdot Q=P^+Q^-+P^-Q^+$ and the mass-shell conditions become
\begin{equation}
2p_n^+p_n^-=(\mA c)^2,
\qquad
2q_n^+q_n^-=(\mB c)^2,
\label{eq:massshell}
\end{equation}
with all four light-cone components $p_n^{\pm}$, $q_n^{\pm}$ positive. The mass shell
is solved once and for all by the rapidity parametrization
\begin{equation}
p_n^{\pm}=\frac{\mA c}{\sqrt2}\,e^{\pm y_n^{(A)}},
\quad
q_n^{\pm}=\frac{\mB c}{\sqrt2}\,e^{\pm y_n^{(B)}},
\quad
a_n=c\tanh y_n^{(A)},
\quad
b_n=c\tanh y_n^{(B)},
\label{eq:rapidity}
\end{equation}
where $y_n^{(A)}$ [$y_n^{(B)}$] is the rapidity of $A$ [$B$] after its
$n$th collision and $a_n$ [$b_n$] the corresponding velocity. The massless
mediator has only a single nonvanishing component,
\begin{equation}
K_n=\eta\,\big[\sqrt2\,|k_n|,\,0\big],
\qquad
K_n'=\eta\,\big[0,\,\sqrt2\,|k_n'|\big],
\end{equation}
in the bracket notation $[P^+,P^-]$. 
This is the technical reason why the problem linearizes. Since each mediator leg
carries a single light-cone component, the plus and the minus projections of
energy--momentum conservation at any collision involve \emph{separate} sectors,
as Eq.~(\ref{eq:conservation}) below makes explicit. The two sectors are coupled
solely through the mass-shell condition (\ref{eq:massshell}), which fixes
$p_n^+$ once $p_n^-$ is known.

\section{Exact solution for the momenta}
\label{sec:solution}

\subsection{Conservation laws in light-cone coordinates}

Each individual collision is essentially one-dimensional Compton scattering --- the
elastic collision of a massless quantum with a free massive particle, in
backscattering geometry --- whose exact kinematics we simply write in
light-cone components~\cite{BLP}. 
Four-momentum conservation at the two collisions reads, for $\eta=+1$ (the case for $\eta=-1$ will be discussed in Sec.~\ref{sec:attractive}),
\begin{equation}
\underbrace{P_{n-1}+K_{n-1}'=P_n+K_n}_{\textstyle A_n},
\qquad
\underbrace{Q_{n-1}+K_n=Q_n+K_n'}_{\textstyle B_n}.
\label{eq:vectorcons}
\end{equation}
At $A_n$, particle $A$ absorbs the left-mover $K_{n-1}'$ emitted at $B_{n-1}$
and emits the right-mover $K_n$; at $B_n$ particle $B$ absorbs that same $K_n$
and emits the left-mover $K_n'$, which travels back to $A_{n+1}$. Projecting
onto the plus and minus components, and using $K_n=[\sqrt2|k_n|,0]$ and
$K_n'=[0,\sqrt2|k_n'|]$, gives
\begin{subequations}
\label{eq:conservation}
\begin{align}
\sqrt2\,|k_n| &= p_{n-1}^+-p_n^+ = q_n^+-q_{n-1}^+,
\\
\sqrt2\,|k_{n-1}'| &= p_n^--p_{n-1}^-,
\qquad
\sqrt2\,|k_n'| = q_{n-1}^--q_n^- .
\end{align}
\end{subequations}
Eliminating the mediator, the sums
\begin{subequations}
\label{eq:conveyor}
\begin{align}
p_n^++q_n^+ &= p_0^++q_0^+ \equiv \frac{\Lambda_1}{\sqrt2},
\\
p_{n+1}^-+q_n^- &= p_0^-+q_0^-+\sqrt2\,|k_0'| \equiv \frac{\Lambda_2}{\sqrt2},
\end{align}
\end{subequations}
are conserved for all $n\ge0$: the plus components of $A$ and $B$ share a
fixed budget, while the minus components share a budget shifted by one
collision --- the mediator is the conveyor that carries the exchanged
component from one particle to the other. The product of the two constants
is a Lorentz invariant, fixed by the total invariant energy $\sqrt{s}$
of the three-body system:
\begin{equation}
\Lambda_1\Lambda_2 = (P_0+Q_0+K_0')^2 \equiv s/c^2 .
\label{eq:sdef}
\end{equation}

\subsection{Solving the recurrence}

Substituting $q_n^+=\Lambda_1/\sqrt2-p_n^+$ and the mass shell
(\ref{eq:massshell}) into Eq.~(\ref{eq:conveyor}b) yields a closed recursion
for the single variable $p_n^-$,
\begin{equation}
p_{n+1}^-=\frac{\Lambda_2}{\sqrt2}
-\frac{(\mB c)^2\,p_n^-}{\sqrt2\,\Lambda_1\,p_n^--(\mA c)^2},
\label{eq:mobius}
\end{equation}
a fractional-linear (M\"obius) map~\cite{Beardon1991}. (A
one-variable map of the same fractional-linear type governs the
mirror-symmetric equal-mass system of Ref.~\cite{ArtigueRel2024} --- the
$\mB\to\infty$ limit of ours; see Sec.~\ref{sec:attractive}.) Writing
\begin{equation}
p_n^-=g_n/f_n
\quad
\textrm{with}
\quad
g_n\equiv(\mA c)^2(f_{n+1}+f_n)/(\sqrt2\Lambda_1)
\end{equation}
linearizes it into the
three-term recursion
\begin{equation}
(\mA c)^2f_{n+2}-\big[s/c^2-(\mA c)^2-(\mB c)^2\big]f_{n+1}+(\mB c)^2f_n=0,
\label{eq:threeterm}
\end{equation}
whose characteristic roots are
\begin{equation}
x_\pm=\frac{\mB}{\mA}\,e^{\pm\theta},
\qquad
\cosh\theta=\alpha\equiv\frac{s-(\mA c^2)^2-(\mB c^2)^2}{2(\mA c^2)(\mB c^2)} .
\label{eq:theta}
\end{equation}
The invariant $\alpha$ plays the role that the mass ratio plays in the
Newtonian problem: there, the analogous recursion has
\emph{complex} roots $e^{\pm i\theta}$ with
$\tan(\theta/2)=\sqrt{m/M}$, and the momenta oscillate~\cite{Ee2012}; here the
roots are \emph{real} and the momenta evolve monotonically --- rotation has
become boost.\footnote{The map is the projective action of a $2\times2$
matrix, with conjugacy invariant $2\alpha$: hyperbolic for $|\alpha|>1$,
elliptic for $|\alpha|<1$, its fixed points being the terminal states
[cf.~Eq.~(\ref{eq:terminal})]. The classification is collected in
Appendix~\ref{app:map}.} Positivity of the mediator energy ($\eta=+1$) guarantees the
hyperbolic regime: using Eqs.~(\ref{eq:conveyor})--(\ref{eq:sdef}) one finds
\begin{equation}
\alpha-\frac{1+\beta^2}{2\beta}
=\eta\,\frac{\sqrt2\,|k_0'|\,(p_0^++q_0^+)}{(\mA c)(\mB c)},
\qquad
\beta\equiv\frac{q_0^+/(\mB c)}{p_0^+/(\mA c)}=e^{\,y_0^{(B)}-y_0^{(A)}},
\label{eq:repulsive}
\end{equation}
so that for $\eta=+1$ we have $\alpha\ge(1+\beta^2)/(2\beta)\ge1$ always, and
$\theta$ is real. (The opposite inequality of the attractive branch opens
the oscillatory window explored in Sec.~\ref{sec:attractive}.)

\subsection{Closed-form momenta}

The general solution of the linear recursion (\ref{eq:threeterm}) is
$f_n=d_+x_+^{\,n}+d_-x_-^{\,n}$, and the initial data fix the ratio
$d_-/d_+=e^{\theta-2\phi}$ --- equivalently the phase $\phi$, given in
closed form by Eq.~(\ref{eq:phidef}) below; the two exponentials then combine into a
single hyperbolic cosine,
$f_n\propto(\mB/\mA)^{n}\cosh[(n-\tfrac12)\theta+\phi]$, which is why every
closed form below is built from such hyperbolic cosines. With the shorthands
\begin{equation}
\mathcal{C}_{n}\equiv\cosh\!\big[\big(n+\tfrac12\big)\theta+\phi\big],
\qquad
\mathcal{B}_n\equiv\mA c\,\mathcal{C}_{n-1}+\mB c\,\mathcal{C}_n,
\label{eq:shorthands}
\end{equation}
the result is
\begin{subequations}
\begin{align}
p_n^- &= \frac{\mA c\,\mathcal{B}_n}{\sqrt2\,\Lambda_1\,\mathcal{C}_{n-1}},
&
p_n^+ &= \frac{\Lambda_1\,\mA c\,\mathcal{C}_{n-1}}{\sqrt2\,\mathcal{B}_n},
\label{eq:pnm}
\\
q_n^- &= \frac{\mB c\,\mathcal{B}_n}{\sqrt2\,\Lambda_1\,\mathcal{C}_n},
&
q_n^+ &= \frac{\Lambda_1\,\mB c\,\mathcal{C}_n}{\sqrt2\,\mathcal{B}_n},
\\
|k_n'| &= \frac{(\mA c)(\mB c)\sinh^2\theta}{2\,\Lambda_1\,\mathcal{C}_{n-1}\mathcal{C}_n},
\label{eq:knpabs}
\\
|k_n| &= \frac{(\mA c)(\mB c)\,\Lambda_1\sinh^2\theta}{2\,\mathcal{B}_{n-1}\mathcal{B}_n},
\label{eq:knabs}
\end{align}
\end{subequations}
where the phase $\phi$ is fixed by the initial condition,
\begin{equation}
e^{2\phi}=\frac{\beta\,e^{\theta}-1}{e^{\theta}-\beta}\,,
\label{eq:phidef}
\end{equation}
which is real precisely in the repulsive regime (\ref{eq:repulsive}),
where $e^{\theta}\ge\max(\beta,\beta^{-1})$; for equal initial
rapidities ($\beta=1$) it gives $\phi=0$ at once. Substituting $f_n$ back
into $p_n^-=g_n/f_n$ gives Eqs.~(\ref{eq:pnm})--(\ref{eq:knabs}); the mass
shell (\ref{eq:massshell}) then provides the plus components, and the
velocities follow from the rapidity parametrization (\ref{eq:rapidity}),
$a_n/c=\tanh y_n^{(A)}$ with $e^{2y_n^{(A)}}=p_n^+/p_n^-$.

The structural parallel with the Newtonian solution of
Ref.~\cite{Ee2012} is now manifest: there, the velocities after the $n$th
collision are $\cos$ and $\sin$ of $n\theta$ shifted by a phase fixed by the
initial condition; here they are $\cosh$ and $\sinh$ of $n\theta+\phi$.
All closed forms of this paper have been verified symbolically and against
exact event-driven simulations of the collision dynamics --- for rest-frame,
boosted, asymmetric, and strong-coupling initial data, and for both signs of
the mediator energy --- to relative accuracy $\lesssim10^{-13}$ in double
precision, and identically in exact symbolic arithmetic.

\subsection{Example: repulsion from rest, and the worldline picture}
\label{sec:example}

\begin{figure}[t]
\centering
\includegraphics[width=0.8\linewidth]{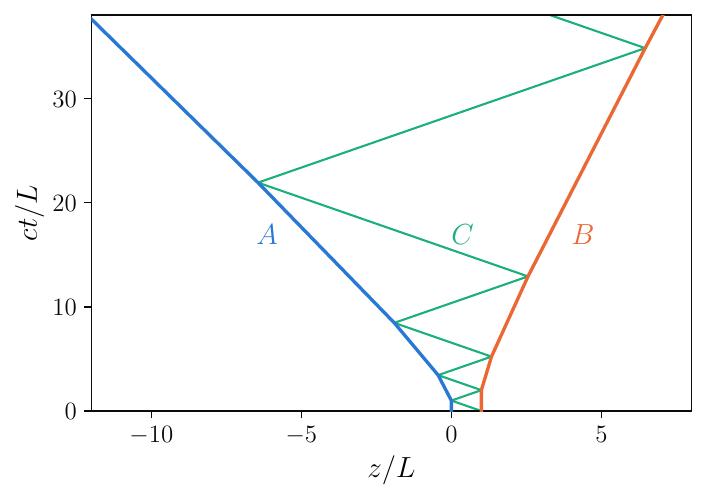}
\caption{\label{fig:worldlines}%
Worldlines of the exact solution in the initial rest frame of $A$ and $B$
($\mB/\mA=1.5$, $|k_0'|=0.1\,\mA c$, initial separation $L$). The massless
mediator $C$ (light trajectory) bounces between $A$ and $B$,
pushing them apart and red-shifting at every reflection. The
collision sequence is infinite, but the momentum transfer decreases
exponentially [Eq.~(\ref{eq:knpabs})], and the velocities of $A$ and $B$
converge to the terminal values of Eq.~(\ref{eq:terminal}).}
\end{figure}

The simplest special case is $A$ and $B$ initially at rest ($p_0=q_0=0$) with
the mediator bouncing between them. Then
\begin{equation}
\Lambda_1=(\mA{+}\mB)c,\quad
\beta=1,\quad \phi=0,\quad
\alpha=1+\frac{|k_0'|(\mA{+}\mB)}{\mA \mB c},
\label{eq:restframe}
\end{equation}
and the closed forms simplify accordingly. Figure~\ref{fig:worldlines} shows
the exact worldlines: the mediator's zigzag is the mechanical skeleton of the
repulsion --- a classical ladder diagram in which every rung is an on-shell
massless exchange.\footnote{In the field-theory ladder the rungs are
spacelike, virtual quanta; here every rung is a null, on-shell segment ---
the resemblance is structural rather than diagrammatic.}

Two features distinguish the relativistic system qualitatively from its
Newtonian counterpart. First, the collision sequence never terminates:
because $C$ moves at the speed of light it always catches the receding
particles, so there is no finite final collision number $N$ as in
Refs.~\cite{Ee2012,Ee2015}. Instead, the exchanged momentum
(\ref{eq:knpabs}) decreases exponentially in $n$,
\begin{equation}
|k_n'|\to
2(\mA c)(\mB c)\sinh^2\theta\,e^{-2\phi}e^{-2n\theta}/\Lambda_1,
\end{equation}
and the mediator
becomes arbitrarily soft --- reminiscent of the infrared softening of the
radiation that accompanies charged-particle scattering~\cite{BlochNordsieck1937}. Second, the
velocities converge exponentially to terminal values obtained from the
$n\to\infty$ limit of Eqs.~(\ref{eq:pnm})--(\ref{eq:knabs}); for the
rest-frame initial condition (\ref{eq:restframe}),
\begin{equation}
e^{-y_\infty^{(A)}}=\frac{\mA+\mB e^{\theta}}{\mA+\mB},
\qquad
e^{+y_\infty^{(B)}}=\frac{(\mA+\mB)\,e^{\theta}}{\mA+\mB e^{\theta}},
\label{eq:terminal}
\end{equation}
with $a_\infty=c\tanh y_\infty^{(A)}<0<b_\infty=c\tanh y_\infty^{(B)}$: the
pair dissociates, leaving the soft mediator bouncing between two receding
particles. [Both relations follow from
$\mathcal{C}_n/\mathcal{C}_{n-1}\to e^{\theta}$ in
Eqs.~(\ref{eq:pnm})--(\ref{eq:knabs}).]

The invariant $\alpha$ of Eq.~(\ref{eq:theta}) has, in fact, a direct
kinematic meaning. Because the mediator softens exponentially, the pair
alone comes to carry the whole invariant: by four-momentum conservation
and $K_n'^{\,2}=0$,
\begin{equation}
(P_n+Q_n)^2=\frac{s}{c^{2}}-2\,(P_n+Q_n)\cdot K_n'
=\frac{s}{c^{2}}-2\,\Lambda_1|k_n'|\;\longrightarrow\;\frac{s}{c^{2}},
\label{eq:pairinv}
\end{equation}
the remainder vanishing exponentially by Eq.~(\ref{eq:knpabs}). For two
free particles, on the other hand,
$\cosh\big(y^{(B)}-y^{(A)}\big)=[(P+Q)^2-(\mA c)^2-(\mB c)^2]
/[2(\mA c)(\mB c)]$, which is precisely Eq.~(\ref{eq:theta}). Hence
$\theta$ is the \emph{terminal relative rapidity} of the receding pair,
$\theta=y_\infty^{(B)}-y_\infty^{(A)}$, so that the invariant which fixes
the whole solution, $\alpha=\cosh\theta$, is the Lorentz factor of the
pair's relative motion once the mediator has been spent. At finite $n$ the same quantity is
exact and monotone: the closed forms give
$y_n^{(B)}-y_n^{(A)}=\ln(\mathcal{C}_n/\mathcal{C}_{n-1})$, which climbs
monotonically to $\theta$, collision by collision.

\section{Collision times and positions}
\label{sec:times}

Between collisions all velocities are constant, so the time intervals
$\Delta_n\equiv t_n'-t_n$ (flight $A_n\to B_n$) and
$\Delta_n'\equiv t_{n+1}-t_n'$ (flight $B_n\to A_{n+1}$) obey the recursions
\begin{equation}
\Delta_n'=\frac{c-a_n}{c+a_n}\,\Delta_n,
\qquad
\Delta_{n+1}=\frac{c+b_n}{c-b_n}\,\Delta_n',
\label{eq:doppler}
\end{equation}
where $a_n$ and $b_n$ are the velocities of $A$ and $B$ after their $n$th
collisions. Equations~(\ref{eq:doppler}) are nothing but the Doppler factors of a
light clock whose mirrors recede. Iterating them with
Eqs.~(\ref{eq:pnm})--(\ref{eq:knabs}) gives, remarkably, perfect-square
ratios,
\begin{equation}
\Delta_n=\Delta_1\,\frac{\mathcal{C}_{n-1}^{2}}{\mathcal{C}_0^{2}},
\qquad
\Delta_n'=\Delta_1\,\frac{\mathcal{B}_n^{2}}{\Lambda_1^{2}\,\mathcal{C}_0^{2}},
\label{eq:Deltan}
\end{equation}
with $c\Delta_1=z_1'-z_1$.
The sums over flight segments telescope,
yielding the collision times $t_n$, $t_n'$ and positions $z_n$, $z_n'$ in
closed form; the results and their derivation are collected in
Appendix~\ref{app:times}. Here we quote the outputs needed for the
force law, the two families of flight lengths. With the initial separation
scale set by $z_1'-z_1$, the legs traversed by the right-movers have lengths
\begin{subequations}
\begin{equation}
z_n'-z_n=(z_1'-z_1)\,\frac{\mathcal{C}_{n-1}^{2}}{\mathcal{C}_0^{2}}.
\label{eq:ynxn}
\end{equation}
The legs
traversed by the left-movers are
\begin{equation}
z_n'-z_{n+1}
=(z_1'-z_1)\,\frac{\mathcal{B}_n^{2}}{\Lambda_1^{2}\,\mathcal{C}_0^{2}} .
\label{eq:ynxnp}
\end{equation}
\end{subequations}
Every flight length is the perfect square of one of the amplitudes
$\mathcal{C}_n$ or $\mathcal{B}_n$ that build the mediator momenta
(\ref{eq:knpabs}) and (\ref{eq:knabs}) --- the structural fact behind the
force law of the next section.

\section{The effective Coulomb potential}
\label{sec:potential}

As in the Newtonian problem~\cite{Ee2012}, total energy conservation invites
us to interpret the mediator energy as the potential energy of the pair,
\begin{equation}
V \;=\; |k'|\,c
\qquad\text{(mediator in flight toward $A$)},
\end{equation}
and to trade the collision index $n$ for a geometric separation of the pair.
Because the mediator moves at the speed of light, the collision coordinates
\emph{are} the flight lengths:
\begin{equation}
z_n'-z_n=c\,\Delta_n,
\qquad
z_n'-z_{n+1}=c\,\Delta_n' ,
\label{eq:legs}
\end{equation}
so the two right-moving legs adjacent to the left-moving leg $n$ have lengths
$z_n'-z_n$ and $z_{n+1}'-z_{n+1}$. The choice of ``the'' separation sampled by
a discrete flight is a convention, and the choice proves consequential.
Define the \emph{geometric-mean separation}
\begin{align}
\tilde r_n&\equiv\sqrt{(z_{n-1}'-z_n)(z_n'-z_{n+1})},
\nonumber\\
\tilde r_n'&\equiv\sqrt{(z_n'-z_n)(z_{n+1}'-z_{n+1})}.
\label{eq:gmdef}
\end{align}
Here $\tilde r_n$ is the geometric mean of the two left-moving legs that
bracket the right-moving leg $n$ (and pairs with $|k_n|$), while
$\tilde r_n'$ is that of the two right-moving legs bracketing the
left-moving leg $n$ (pairing with $|k_n'|$).
At $n=1$ the first geometric mean involves the leg $z_0'-z_1$: here the event $B_0$
denotes the virtual emission event of the initial left-mover --- the
intersection of its backward extension with the worldline of $B$ prior to
$B_1$ --- so that $z_0'$ is fixed by the initial data and coincides with the
$n=0$ case of Eq.~(\ref{eq:ynxnp}).
Then the central result of this paper is an \emph{exact discrete Coulomb
law}: for every $n$,
\begin{equation}
\boxed{\;
|k_n|\;\tilde r_n
\;=\;
|k_n'|\;\tilde r_n'
\;=\;J\;=\;\frac{|k_0'|\,(z_1'-z_1)}{\beta}\;,
\;}
\label{eq:exactJ}
\end{equation}
with $\beta$ the initial-condition invariant of Eq.~(\ref{eq:repulsive}):
the momentum of each leg of the mediator, multiplied by the geometric mean
of the lengths of the two opposite-moving legs that bracket it, is a
constant of the motion. Equivalently, $V(\tilde r)=Jc/\tilde r$ holds
\emph{exactly at every collision} --- not merely asymptotically or in a
continuum limit --- and the effective force between $A$ and $B$, defined as
$F=-dV/d\tilde r$, is the inverse-square repulsion
$F=Jc/\tilde r^{\,2}$.\footnote{The instantaneous momentum-transfer rate
coincides with $-dV/d\tilde r$ only in the adiabatic regime: at finite
$\theta$ the mediator retains part of the exchanged momentum, so $A$ and
$B$ do not receive equal and opposite impulses.} The proof is immediate ---
this is what the perfect squares (\ref{eq:ynxn}) and (\ref{eq:ynxnp}) are
for. The geometric means (\ref{eq:gmdef}) evaluate to
\begin{equation}
\tilde r_n'=(z_1'-z_1)\,
\frac{\mathcal{C}_{n-1}\mathcal{C}_n}{\mathcal{C}_0^{2}},
\qquad
\tilde r_n=(z_1'-z_1)\,
\frac{\mathcal{B}_{n-1}\mathcal{B}_n}{\Lambda_1^{2}\,\mathcal{C}_0^{2}},
\label{eq:gmvalues}
\end{equation}
and their products with the mediator momenta (\ref{eq:knpabs}) and
(\ref{eq:knabs}) cancel the $n$~dependence on sight:
\begin{equation}
|k_n'|\,\tilde r_n'
=|k_n|\,\tilde r_n
=\frac{(\mA c)(\mB c)\sinh^{2}\theta\,(z_1'-z_1)}
{2\,\Lambda_1\,\mathcal{C}_0^{2}} .
\label{eq:Jraw}
\end{equation}
Equation~(\ref{eq:phidef}) yields
$\mathcal{C}_0^{2}=\cosh^{2}[\tfrac{\theta}{2}+\phi]
=\beta^{2}(\alpha^{2}-1)/(2\alpha\beta-1-\beta^{2})$, and
Eq.~(\ref{eq:repulsive}) in the form
$(2\alpha\beta-1-\beta^{2})/(2\beta)=|k_0'|\,\Lambda_1/[(\mA c)(\mB c)]$
collapses the constant to $J=|k_0'|(z_1'-z_1)/\beta$.

The geometric mean is essential to the exactness: sampling with either
single adjacent leg, or with their arithmetic mean, fails by an exact,
monotonically drifting factor [Appendix~\ref{app:area},
Eq.~(\ref{eq:sampling})]. Equation~(\ref{eq:exactJ}) is therefore best
read as an exact first integral of the collision map --- of action form,
Eq.~(\ref{eq:invariant}) --- whose sampling-independent content is the
adiabatic law $J=kr$ of Sec.~\ref{sec:unified}; the same appendix
exhibits the reflection--dilation pairing that makes the geometric mean,
and only it, exact.

An invariant of this kind was proven, for the
mirror-symmetric configuration, by Artigue~\cite{ArtigueRel2024}: there the
product of the separation with the mediator energy, divided by $E+pc$ of the
massive particle, is constant from collision to collision, and
(mediator energy)$\,\times\,$(separation) converges as $t\to\pm\infty$.
Equation~(\ref{eq:exactJ}) contains that statement as its $\mB\to\infty$
limit (Sec.~\ref{sec:attractive}) and generalizes it to arbitrary masses and
to both signs of the mediator energy, in the form of a pure action: the
geometric-mean sampling replaces the auxiliary factor $E+pc$ by the
symmetric pairing of the two legs that bracket each mediator flight.
In exact event-driven simulations the law holds to machine precision;
Fig.~\ref{fig:phasespace}(b) displays it geometrically.\footnote{Exactness of this kind is not
automatic, and the Newtonian case is instructive. In the block-and-wall
problem of Ref.~\cite{Ee2012} the product (ball momentum)$\,\times\,$(block
distance) is \emph{exactly} conserved provided it is evaluated at the
collisions with the \emph{fixed wall} --- a consequence of Galperin's
unfolding, in which the reduced motion is rectilinear and both collision
lines pass through the apex, so that the angular momentum about the apex is
conserved; evaluated at the block--ball collisions the same product drifts
by $O(m/M)$ per collision~\cite{SkinnerNeishtadt2024}. Once \emph{both}
partners recoil, as in Ref.~\cite{Ee2015}, none of the natural
samplings --- either collision, or their geometric mean --- is exact.
What
the relativistic massless mediator restores is exactness for the genuine
two-body problem, with the geometric mean (\ref{eq:gmdef}) playing the role
that the fixed wall plays in the Newtonian case: every flight length is the
perfect square of one of the amplitudes $\mathcal{C}_{n-1}$ and
$\mathcal{C}_n$ whose product forms the denominator of $|k_n'|$, so the geometric mean cancels it
identically --- a gift of the multiplicative Doppler factors of
Sec.~\ref{sec:times}.}

The coupling $J$ has two properties worth noting. First, it has the
dimension of \emph{action}, and it is Lorentz invariant: the numerator obeys
\begin{equation}
|k_0'|\,(z_1'-z_1)=\tfrac12\,K_0'\cdot(Z_1'-Z_1),
\label{eq:invariant}
\end{equation}
where $Z_1$ and $Z_1'$ are the four-positions of
the first two collisions $A_1$ and $B_1$ [the leg $A_1\to B_1$ is lightlike,
$c(t_1'-t_1)=z_1'-z_1$], while
$\beta$, by Eq.~(\ref{eq:repulsive}), is the exponential of a rapidity
\emph{difference}, invariant by itself; hence $J$ is a Lorentz scalar with
the dimension of action. Second, this is precisely the structure of the
long-range couplings of nature: $e_Ae_B\,e^{2}/(4\pi\epsilon_{0})$ of
electrostatics and $G\mA\mB$ of Newtonian gravity both carry the dimension
of action$\,\times\,c$, and a $1/r$ potential is the natural one for a
scale-free mediator to support with such a coupling. We do not attempt to
interpret the analog of the fine-structure constant --- $J$ is set by the
initial data, not by nature --- but the dimensional mechanism is the same.
A $1/r$ potential needs a coupling of dimension action$\,\times\,$velocity.
A massive mediator would bring a scale of its own, $mc^{2}$; the
massless mediator brings none, so the only Lorentz-invariant scale it can
transmit is the action
(\ref{eq:invariant}) handed to it by the initial data.
Dimensional analysis alone cannot forbid corrections by functions of the
dimensionless invariants of the system (such as $\alpha$ and $\beta$); the
content of Eq.~(\ref{eq:exactJ}) is that, for the geometric-mean
separation, such corrections are absent identically.

\section{The attractive branch: negative-energy mediator and a mechanical
atom}
\label{sec:attractive}

The closed forms of Secs.~\ref{sec:solution}--\ref{sec:potential} continue
analytically to the branch $\eta=-1$, in which the mediator carries negative
energy and every collision pulls $A$ and $B$ \emph{together}. Such an object
cannot arise from elastic collisions of physical particles --- which is why
we call the branch formal --- but mediators of this kind have been
introduced and studied by Artigue as ``toy gravitons,'' first in Newtonian
mechanics, where a light particle of \emph{negative mass} between two
positive masses produces attraction~\cite{Artigue:2018jwu}, and recently in
the relativistic one-dimensional setting, where the mediator is massless
with negative energy~\cite{ArtigueRel2024};
following that precedent, we treat the branch on the same footing as the
repulsive one.

\begin{figure}[t]
\centering
\includegraphics[width=0.8\linewidth]{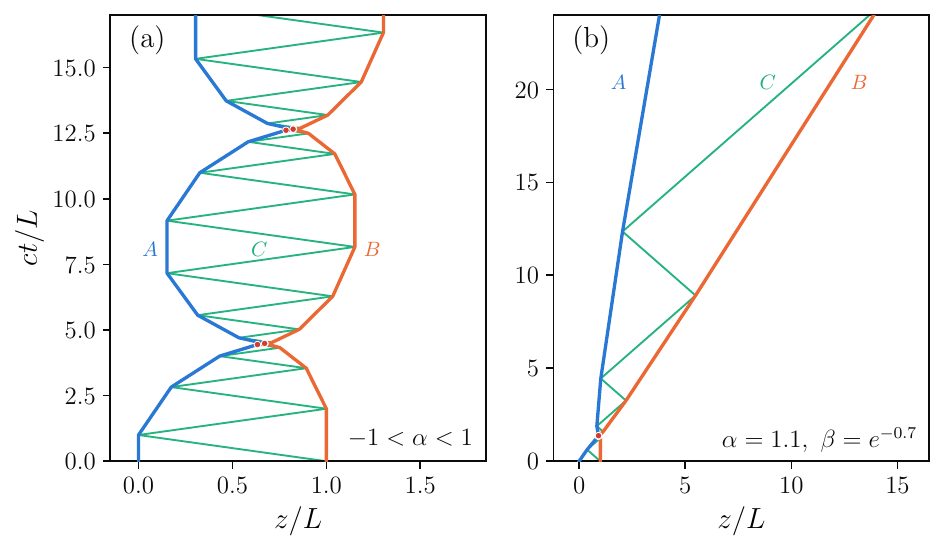}
\caption{\label{fig:atom}%
Worldlines of the attractive branch ($\eta=-1$), from the exact
event-driven collision algebra (axes scaled independently; every mediator
segment is lightlike). Green: ordinary negative-energy mediator legs;
red: the isolated sign-flipped legs at the innermost approach, produced
by the collisions violating the threshold (\ref{eq:threshold}) (see
text). (a)~Bound window: $\mB/\mA=1.5$, blocks initially at rest,
$|k_0'|=(1-\cos\tfrac{\pi}{8})\,\mA\mB c/(\mA+\mB)$
($\alpha=\cos\vartheta$, $\vartheta=\pi/8$) --- a one-dimensional atom,
periodic with eight collision pairs (sixteen collisions) per cycle;
per cycle exactly one core exchange turns the pair around. (b)~Unbound regime: $\mB=\mA$,
$\alpha=1.10$, $\beta=e^{-0.7}<1$ --- the pair approaches, crosses the
core through one such exchange, and dissociates.}
\end{figure}

For $\eta=-1$ the constant $\Lambda_2$ of Eq.~(\ref{eq:conveyor}) becomes
$\Lambda_2=\sqrt2\,(p_0^-+q_0^--\sqrt2|k_0'|)$, and Eq.~(\ref{eq:repulsive})
reverses: $\alpha<(1+\beta^2)/(2\beta)$. Four regimes open up, classified
by the total invariant mass (Fig.~\ref{fig:atom}):
\begin{itemize}
\item $\alpha\le-1$, i.e. $\sqrt s\le|\mA-\mB|c^2$: reachable for
$\eta=-1$ at large enough $|k_0'|$, and at larger $|k_0'|$ still $s$ itself
turns negative, the total four-momentum becoming spacelike. The M\"obius
map (\ref{eq:mobius}) is
hyperbolic again, but the continuation has left the physical region
altogether: the lighter particle is a negative-energy state at \emph{every}
collision, not only at the isolated turning-point collisions of the bound
window below. We do not pursue this branch.
\item $-1<\alpha<1$, i.e.
\begin{equation}
|\mA-\mB|\,c^2\;<\;\sqrt{s}\;<\;(\mA+\mB)\,c^2 :
\label{eq:massdefect}
\end{equation}
the three-body invariant mass lies \emph{below} the sum of the constituent
masses --- the familiar sub-threshold condition for binding --- and the
pair is \emph{bound}. The
parameters $\theta$ and $\phi$ become pure imaginary,
$\theta=i\vartheta$, $\phi=i\tilde\varphi$ with $\cos\vartheta=\alpha$, and
every hyperbolic function in Eqs.~(\ref{eq:pnm})--(\ref{eq:knabs}) and
(\ref{eq:Deltan})--(\ref{eq:ynxnp}) becomes trigonometric. The momenta
oscillate, the phase $\chi_n\equiv n\vartheta+\tilde\varphi$ advancing by $\vartheta$
per collision pair, and so does the separation,
\begin{equation}
z_n'-z_n=(z_1'-z_1)\,
\frac{\cos^{2}[(n-\tfrac12)\vartheta+\tilde\varphi]}
{\cos^{2}[\tfrac{\vartheta}{2}+\tilde\varphi]},
\end{equation}
[the companion legs~(\ref{eq:ynxnp}) turn trigonometric in the same way],
and the collision sequence never terminates: a purely mechanical
one-dimensional \emph{atom}, held together by the exchange of a single
massless quantum, with the weak-coupling potential $V(r)=-Jc/r$ [the
$\eta=-1$ continuation of the exact law (\ref{eq:exactJ})]. On this
branch Eq.~(\ref{eq:exactJ}) holds in magnitude at every collision;
equivalently the product (mediator energy)$\,\times\,\tilde r/c$, equal
to $-J$ on the ordinary negative-energy legs [the potential
$V=-Jc/\tilde r$ above], flips to $+J$ precisely on the isolated
sign-flipped legs at the innermost approach, described below. The
hyperbolic--trigonometric switch is the same one that separates scattering
from bound motion in the Kepler problem~\cite{LandauMechanics}; here it
is controlled by the
invariant mass through Eq.~(\ref{eq:massdefect}).
One qualification, visible in Fig.~\ref{fig:atom}(a), is required. An
elastic collision with a negative-energy mediator is well defined only
below a threshold: at $A_n$ the incoming left-mover deposits its entire
(now negative) minus component, $p_n^-=p_{n-1}^--\sqrt2\,|k_{n-1}'|$, so
the outgoing $A$ is a positive-energy particle only if
\begin{equation}
2\,|k_{n-1}'| \;<\; \mA c\,e^{-y_{n-1}^{(A)}},
\qquad\text{i.e.}\qquad
2\,|k_{n-1}'|\,c \;<\; E_A-p_Ac,
\label{eq:threshold}
\end{equation}
where $E_A$ and $p_A$ are the energy and momentum of $A$ just before the
collision [at $B_n$, similarly, $2|k_n|<\mB c\,e^{+y_{n-1}^{(B)}}$].
Written with the \emph{signed} minus-component of the incoming leg the
condition is simply $p_n^->0$; Eq.~(\ref{eq:threshold}) is its form for a
leg carrying the branch's nominal, negative-energy sign; at a collision
whose incoming leg has itself flipped sign --- the situation immediately
after a violation --- it is met automatically. The repulsive branch,
which \emph{adds} the corresponding components, has no such threshold. In the attractive branch the approach violates
Eq.~(\ref{eq:threshold}) after finitely many collisions, on two counts at
once: the mediator blueshifts, $|k'|=J/\tilde r$, while the absorber
accelerates toward its partner and its light-front budget
$\mA c\,e^{-y^{(A)}}$ is exponentially depleted. What happens then is
dictated by the same energy--momentum conservation that governs every
other collision, and it admits only one solution: the absorber's new
minus component, $p^-\!\to p^--\sqrt2\,|k'|$, is now negative, and the
mass shell $2p^+p^-=(\mA c)^{2}$ then forces both components negative
--- conservation itself places the absorber (say $A$) on the
negative-energy root, and the leg it emits, $k^+=p^+_{\rm in}-p^+_{\rm out}>0$, carries
\emph{positive} energy and outward momentum. This flipped leg is the
inner bounce: absorbed by $B$, it kicks $B$ \emph{away} rather than
inward, and the flipped left-mover that $B$ returns does the same for
$A$ while restoring its minus component to a positive value --- $A$
rejoins the positive-energy shell, emits an ordinary negative-energy
leg, and the pulls resume, now decelerating the receding pair. We call
these isolated sign-flipped collisions at the turning point the
\emph{core exchanges}; per cycle, exactly one massive particle dips at
exactly one collision, and exactly two mediator legs briefly carry
positive energy [the ``tachyonic'' collisions identified by
Artigue~\cite{ArtigueRel2024}; in the bound window they are unavoidable,
since a pull alone cannot supply an inner turning point]. Which particle
dips, and how the exchanges are distributed over a period, follows from
the signs of the continued amplitudes $\mathcal{B}_n$ and
$\mathcal{C}_n$. The mechanical atom is thus exact as an analytic
continuation, with $A$ and $B$ ordinary positive-energy states
everywhere except at the isolated core exchanges that turn the pair
around.
\item $1\le\alpha<(1+\beta^2)/(2\beta)$: an unbound attractive regime
whose orbit has a single closest approach, crossed through one tachyonic
core exchange. Initial data with $\beta>1$ start \emph{after} the closest
approach: the pair simply dissociates, every collision an ordinary
elastic one. Initial data with $\beta<1$ start \emph{before} it: the pair
approaches, crosses the core, and dissociates [Fig.~\ref{fig:atom}(b)].
\item $\alpha\ge(1+\beta^2)/(2\beta)$: recovered only for $\eta=+1$ --- the
repulsive scattering of Secs.~\ref{sec:solution}--\ref{sec:potential}.
\end{itemize}

The bound regime makes contact with Ref.~\cite{ArtigueRel2024}, and
the relation is in fact an exact limit. The mirror symmetry of the four-particle
configuration studied there makes its central mediator collision a perfect
reflection, so that the dynamics reduces to a single massive particle
exchanging a mediator with a \emph{fixed wall} --- kinematically the
$\mB\to\infty$ limit of the present three-body system, in the frame where
the heavy particle is at rest. Our closed forms stay finite in that limit
and collapse to a one-variable map: with $\sigma_n\equiv E_A+p_Ac$
$=\mA c^{2}e^{y_n^{(A)}}$ evaluated just after the $n$th collision
$A_n$,
Eqs.~(\ref{eq:massshell})--(\ref{eq:conveyor}) give
\begin{equation}
\sigma_n=\frac{(\mA c^{2})^{2}}{2\mathcal{E}-\sigma_{n-1}},
\qquad
\mathcal{E}\equiv E_A+\eta|k|c=\alpha\,\mA c^{2}=\textrm{const},
\label{eq:wallmap}
\end{equation}
the wall absorbing momentum but no energy, $|k_n|=|k_n'|$ [the $B$-side
companion of Eq.~(\ref{eq:reflexact}) with $y^{(B)}\equiv0$]. This is
precisely the fractional-linear map of Ref.~\cite{ArtigueRel2024}, whose
parameters are here
\begin{equation}
\mu=(\mA c^{2})^{2},
\qquad
\Delta\equiv\mathcal{E}^{2}-\mu=(\mA c^{2})^{2}\sinh^{2}\theta ,
\label{eq:walldict}
\end{equation}
so that its discriminant reproduces our trichotomy $|\alpha|>1$,
$|\alpha|=1$, $|\alpha|<1$ as the hyperbolic, parabolic and elliptic
cases~\cite{Beardon1991} of the M\"obius map (\ref{eq:mobius}); the
per-collision invariant proven there ---
(separation)$\,\times\,$(mediator energy)$/\sigma$ --- is our law
(\ref{eq:exactJ}), because in this limit the geometric mean
(\ref{eq:gmdef}) collapses to $\tilde r_n=r_n\,\mA c^{2}/\sigma_n$; and its
criterion for ``tachyonic'' collisions is our threshold
(\ref{eq:threshold}). What the general closed forms add is the recoil of
both partners at finite $\mB$ --- a genuine two-body problem rather than a
particle in a fixed cavity ---, arbitrary initial data, the physical
positive-energy branch, the invariant in the pure action form
(momentum)$\,\times\,$(separation) with no auxiliary weight, and, in the same
closed forms, the collision times and positions. In this precise sense
the closed forms complete the solution of the one-dimensional problem of
two masses exchanging a single massless mediator: arbitrary masses,
arbitrary initial data, and both signs of the mediator energy, with
momenta, collision times, and positions in closed form and the invariant
(\ref{eq:exactJ}) exact at every collision.

\section{Why \texorpdfstring{$1/r$}{1/r}: dispersion, adiabatic invariance, and the limits of the
analogy}
\label{sec:unified}

\subsection{One invariant, two force laws}

The results of Ref.~\cite{Ee2012} and of this paper are two instances
of a single mechanism. When the mediator is light and fast compared to the
massive particles, many collisions occur while the separation $r$ changes
appreciably, and the bouncing mediator is a fast subsystem enclosed by slowly
moving turning points --- precisely the setting of an adiabatic
invariant~\cite{LandauMechanics}.
The bouncing-ball realization of this setting is
classic~\cite{Fermi1949,Ulam1961}: the same product $kr$ appears
there, and the conditions for its persistence --- and for its breakdown
into Fermi acceleration --- were first studied in that context.
For walls held \emph{fixed} at separation $r$, the mediator's orbit in the $(z,p)$ plane is the rectangle $z_A\le z\le z_B$, $p=\pm k$, of area
\begin{equation}
\oint p\,dz = 2\,k\,r,
\label{eq:adiabatic}
\end{equation}
where $k=|p|$ is the magnitude
of the mediator momentum; the adiabatic theorem states that it is this
area --- the action $kr$ --- that survives a deformation of the enclosure
slow on the scale of the bounce period. For the massless mediator the
collision map carries an \emph{exact} counterpart of this invariant,
collision by collision: Eq.~(\ref{eq:exactJ}) states $J=k\,\tilde r$ at
every $n$, with the geometric-mean separation $\tilde r$ in place of $r$.
The two coincide in the adiabatic limit, $\tilde r\to r$ as $\theta\to0$,
so that $J$ is the exact finite-$\theta$ deformation of the adiabatic
action. Appendix~\ref{app:area}
exhibits the surviving area directly: each reflection redshifts the
mediator by exactly the inverse of the factor by which the geometric-mean
width dilates [Eqs.~(\ref{eq:reflexact})--(\ref{eq:gmratio})], so the
geometric-mean rectangle of height $|k|$ and width $\tilde r$ deforms from
tall-and-narrow to short-and-wide at constant area
[Fig.~\ref{fig:phasespace}(b)]. The potential energy of the
pair is then the mediator energy evaluated on the invariant,
\begin{equation}
\boxed{\;
V(r)\;=\;E_C\Big(p=\frac{J}{r}\Big),
\;}
\label{eq:master}
\end{equation}
i.e., \emph{the interaction potential is the dispersion relation of the mediator:}
\begin{center}
\begin{tabular}{@{}lccc@{}}
mediator & $E_C(p)$ & $V$ & $F$\\[1pt]
\hline\\[-8pt]
Newtonian~\cite{Ee2012,Ee2015} & $p^2/2m$ & $1/r^{2}$ & $1/r^{3}$\\[2pt]
massless (this work) & $pc$ & $1/r$ & $1/r^{2}$
\end{tabular}
\end{center}
The rule covers the attractive Newtonian case as well: an elastic collision
reverses the relative velocity irrespective of the sign of the mass, so
$kr$ remains the adiabatic invariant, and for a mediator of negative mass,
$E_C=p^{2}/2m$ with $m<0$, Eq.~(\ref{eq:master}) gives
$V=J^{2}/(2mr^{2})<0$ --- an attractive potential $\propto1/r^{2}$, that is
a force $\propto1/r^{3}$, as found for that system in
Ref.~\cite{Artigue:2018jwu}.
The same two scalings are familiar from elementary quantum mechanics as the
box-size dependence of a confined particle's energy ---
$E\sim\hbar^{2}/(2mL^{2})$ for a massive particle versus
$E\sim\pi\hbar c/L$ for a massless one --- with the action $J$ playing
the role of $\hbar$.
In statistical-mechanical language, Eq.~(\ref{eq:master}) is the adiabat of a
one-molecule gas~\cite{GruberLesne2006}: $F\propto1/r^{3}$ is the adiabat $PL^{3}=\text{const}$ of
the one-dimensional nonrelativistic ideal gas ($\gamma=3$), and
$F\propto1/r^{2}$ is the adiabat $PL^{2}=\text{const}$ of the
one-dimensional photon gas ($\gamma=2$)~\cite{Ehrenfest1916}.

\subsection{What the analogy does and does not say}

A classical particle-exchange model of long-range forces has been studied
before by Lancaster, McGuire, and Titus~\cite{Lancaster:2015lfa}, who let two
heavy particles exchange light mediators constrained to travel always at the
same speed $c$. That constraint --- momentum is handed over but the
mediator's speed is reset by hand --- freezes the momentum transfer per
bounce, and the resulting repulsion falls off only as $F\propto1/r$
(a logarithmic potential). The comparison isolates exactly what relativity
adds here: in the fully elastic treatment the mediator red-shifts at every
reflection, its momentum obeys the adiabatic law $k=J/r$, and the
force steepens from $1/r$ to the true Coulomb $1/r^{2}$. Their
observation that classical exchange of positive-energy particles cannot
produce attraction holds here unchanged, and is what makes
Sec.~\ref{sec:attractive} formal: attraction requires the
negative-energy branch. That mechanical
exchange pictures cannot, in general, reproduce the content of a Feynman
diagram --- the same diagram serving attraction and repulsion --- is a
standing objection to the popular ``ball-throwing''
analogy~\cite{Passon2018}; the model here is not offered as a repair of
that analogy.

It is tempting to read Eq.~(\ref{eq:coulomb-intro}) as ``the photon exchange
force derived mechanically,'' and the temptation should be resisted. 
In quantum field theory the Coulomb potential arises from the
exchange of \emph{virtual} massless quanta, and its $1/r$ form in $3+1$
dimensions reflects the massless propagator $1/\mathbf{q}^{2}$, $\mathbf{q}$ being the
spatial momentum transfer --- indeed, in $1+1$ dimensions the Gauss law
gives a linear potential rather than $1/r$~\cite{Schwinger1962,Coleman1976}.
The mechanism here is different: repeated \emph{on-shell}
exchange whose adiabatic invariant is an action. What the two settings share
is a dimensional argument. A massless mediator introduces no scale of its
own; if the coupling it transmits is an invariant \emph{action} $J$ --- as
Eq.~(\ref{eq:invariant}) shows for the mechanical system, and as
$e_Ae_B\,e^{2}/(4\pi\epsilon_{0}c)$ does for electrostatics --- then $V=Jc/r$ is the
natural outcome: dimensional analysis motivates the form, and the exact
computation, with the geometric-mean sampling (\ref{eq:gmdef}),
establishes it with no residual dependence on the dimensionless
invariants. The mechanical system realizes, with Newton's-cradle
ingredients, the kinematic core of the field-theory folklore, in the
restricted sense established above: \emph{a massless mediator transmitting
an invariant action supports an inverse-square force} --- and here it does
so exactly, collision by collision.

\section{Conclusion}
\label{sec:summary}

We have solved, in closed form, the one-dimensional relativistic three-body
system in which two massive particles interact solely through elastic
collisions with a massless mediator. The light-cone recurrences linearize
just as their Newtonian counterparts do, with hyperbolic functions of the
collision index replacing trigonometric ones; the collision times and
positions follow by telescoping sums. The mediator energy, read as the
potential energy of the pair, obeys the exact discrete Coulomb law
(\ref{eq:exactJ}): $V(\tilde r)=Jc/\tilde r$ at every collision, with the
geometric-mean separation $\tilde r$ and a Lorentz-invariant coupling
$J=|k_0'|(z_1'{-}z_1)/\beta$ of the dimension of action --- the structure
shared by the couplings of electrostatics and gravity. Together with
Ref.~\cite{Ee2012}, the
pair of results is unified by the adiabatic invariant $J=pr$: the mediated
interaction potential is the mediator's dispersion relation sampled at $p=J/r$. A
massless mediator --- available only in relativity --- thereby transmits the
long-range inverse-square repulsion, while the massive Newtonian mediator
transmits the short-range $1/r^{3}$ force. Continued to a formally
negative-energy mediator (the relativistic toy
graviton~\cite{Artigue:2018jwu,ArtigueRel2024}), the same closed forms turn
trigonometric and bind the pair into a one-dimensional mechanical atom,
bound exactly in the sub-threshold window
$|\mA-\mB|c^2<\sqrt{s}<(\mA+\mB)c^{2}$. The
system extends the
block-and-ball family, whose collision counting popularized by
3Blue1Brown~\cite{3b1b_1,3b1b_2} computes $\pi$ and whose quantum version performs
Grover's search~\cite{Brown:2019jvs}, into the relativistic domain, where it
performs yet another party trick: it builds Coulomb's law out of collisions.

Several extensions suggest themselves, and the most immediate is the
\emph{massive} relativistic mediator. Inserting its dispersion relation
$E_C(p)=\sqrt{(mc^2)^2+(pc)^2}$ into the adiabatic estimate
(\ref{eq:master}) gives $V(r)\simeq\sqrt{(mc^2)^2+(Jc/r)^2}$: Coulomb-like
where $Jc/r\gg mc^2$ and, after subtracting the rest energy, dipole-like
($\propto1/r^{2}$) in the opposite regime, with the crossover at the
Compton-like scale $r_*=J/(mc)$ --- a power-law counterpart of the Yukawa
mechanism. The exact solution of that case, in which the hyperbolic
functions of this paper become elliptic and the collision map becomes a
translation on an elliptic curve, will be presented in a companion
paper~\cite{inprep}.

\begin{acknowledgments}
As members of the Korea Pragmatist Organization for Physics Education
(\textsl{KPOP}$\mathscr{E}$), the authors thank the remaining members of
\textsl{KPOP}$\mathscr{E}$ for useful discussions.
This work was supported in part by the National Research Foundation
of Korea (NRF) under Grant Nos. RS-2025-24222969 (J. L.) and
RS-2025-24533579 (U-R. K.).
\end{acknowledgments}

\appendix

\section{Collision times and positions}
\label{app:times}

The Doppler recursions (\ref{eq:doppler}) evaluate to
\begin{align}
\frac{c-a_n}{c+a_n}&=\Big(\frac{\sqrt2\,p_n^-}{\mA c}\Big)^{2}
=e^{-2y_n^{(A)}},
&
\frac{c+b_n}{c-b_n}&=\Big(\frac{\sqrt2\,q_n^-}{\mB c}\Big)^{-2}
=e^{+2y_n^{(B)}},
\end{align}
whence the perfect-square ratios, Eq.~(\ref{eq:Deltan}), and, through
Eq.~(\ref{eq:legs}), the companion flight lengths (\ref{eq:ynxnp}).
Between collisions the
null coordinates telescope: along every right-moving leg
$z_n'-ct_n'=z_n-ct_n$, along every left-moving leg
$z_{n+1}+ct_{n+1}=z_n'+ct_n'$, so that $z_n+ct_n$ grows by $2c\Delta_k$ per
cycle while $z_n-ct_n$ shrinks by $2c\Delta_k'$. All collision times and
positions therefore follow from the two telescoped sums
\begin{subequations}
\label{eq:sums}
\begin{align}
\Sigma_n&\equiv c\!\sum_{k=1}^{n-1}\Delta_k
=\frac{z_1'-z_1}{2\,\mathcal{C}_0^{2}}
\bigg[(n-1)
+\frac{\sinh[(2n{-}2)\theta{+}2\phi]-\sinh2\phi}{2\sinh\theta}\bigg],
\\
\Sigma_n'&\equiv c\!\sum_{k=1}^{n-1}\Delta_k'
=\frac{z_1'-z_1}{2\,\mathcal{C}_0^{2}\Lambda_1^{2}}
\bigg[(n-1)\,s/c^{2}
+\frac{\mathcal{S}_n-\mathcal{S}_1}{2\sinh\theta}\bigg],
\end{align}
\end{subequations}
where $\mathcal{S}_n$ is the sinh companion of the expanded denominator
of Eq.~(\ref{eq:knabs}),
\begin{align}
2\,\mathcal{B}_{n-1}\mathcal{B}_n
&=(s/c^{2})\cosh\theta+(\mA c)^2\cosh[(2n{-}2)\theta{+}2\phi]
\nonumber\\
&\quad+2(\mA c)(\mB c)\cosh[(2n{-}1)\theta{+}2\phi]
+(\mB c)^2\cosh[2n\theta{+}2\phi],
\label{eq:BBexp}
\\
\mathcal{S}_n&\equiv(\mA c)^2\sinh[(2n{-}2)\theta{+}2\phi]
+2(\mA c)(\mB c)\sinh[(2n{-}1)\theta{+}2\phi]
\nonumber\\
&\quad+(\mB c)^2\sinh[2n\theta{+}2\phi],
\end{align}
namely, for all $n\ge1$,
\begin{subequations}
\label{eq:txclosed}
\begin{align}
c\,t_n&=c\,t_1+\Sigma_n+\Sigma_n',
&
z_n&=z_1+\Sigma_n-\Sigma_n',
\\
c\,t_n'&=c\,t_n+c\,\Delta_n,
&
z_n'&=z_n+c\,\Delta_n,
\end{align}
\end{subequations}
with $c\Delta_n=(z_1'-z_1)\mathcal{C}_{n-1}^{2}/\mathcal{C}_0^{2}$
from Eq.~(\ref{eq:Deltan}); Eq.~(\ref{eq:ynxn}) is the difference
$z_n'-z_n$.

\section{The surviving phase-space area}
\label{app:area}

For walls held fixed at separation $r$, the phase-space orbit of the
bouncing mediator is the rectangle $z_A\le z\le z_B$, $p=\pm k$, of area
$\oint p\,dz=2kr$; the adiabatic
theorem~\cite{LandauMechanics,Ehrenfest1916} asserts that this area
survives deformations of the enclosure that are slow on the scale of the
bounce period. In the present model the walls recoil and are not slow,
yet the area survives exactly. This appendix exhibits the mechanism, and
in doing so reproves the exact law (\ref{eq:exactJ}) without the closed
forms of Sec.~\ref{sec:solution}.

At $A_n$, the conservation laws (\ref{eq:conservation}) and the mass
shells (\ref{eq:massshell}) give $\sqrt2\,|k_n|=p_{n-1}^{+}-p_n^{+}$
$=\sqrt2\,p_{n-1}^{+}|k_{n-1}'|/p_n^{-}$, i.e.
\begin{equation}
\frac{|k_n|}{|k_{n-1}'|}
=\frac{p_{n-1}^{+}}{p_n^{-}}
=e^{\,y_{n-1}^{(A)}+y_n^{(A)}}
=\sqrt{\frac{c+a_{n-1}}{c-a_{n-1}}\;\frac{c+a_n}{c-a_n}}\;,
\label{eq:reflexact}
\end{equation}
and, by the same algebra with $+\leftrightarrow-$,
$|k_n'|/|k_n|=e^{-(y_{n-1}^{(B)}+y_n^{(B)})}$ at $B_n$. For a mirror of
infinite mass ($y_{n-1}=y_n$, constant velocity $a$) Eq.~(\ref{eq:reflexact}) is the familiar
Doppler factor $(c+a)/(c-a)$ of Eq.~(\ref{eq:doppler}); for a recoiling
particle, the exact reflection factor is the \emph{geometric mean} of the
frozen-mirror factors taken just before and just after the collision. In
this form the single-collision statement is not new: it is the
one-dimensional version of the classic reading of the Compton shift as a
double Doppler shift, in which the recoiling particle acts as a static
mirror in the frame moving with the mean
rapidity~\cite{NielsenOlsen1966,Kidd1985}. What the present section adds is
its iteration --- the pairing of that factor, collision after collision,
with the dilation of the null legs, which is what turns an approximate
adiabatic invariant into an exact one. (The linearization of a sequence of
null legs reflected by moving mirrors is also the structure underlying
Moore's ray map for a cavity with a prescribed moving
boundary~\cite{Moore1970}; here the boundaries are dynamical and recoil.)

\begin{figure}[t]
\centering
\includegraphics[width=\linewidth]{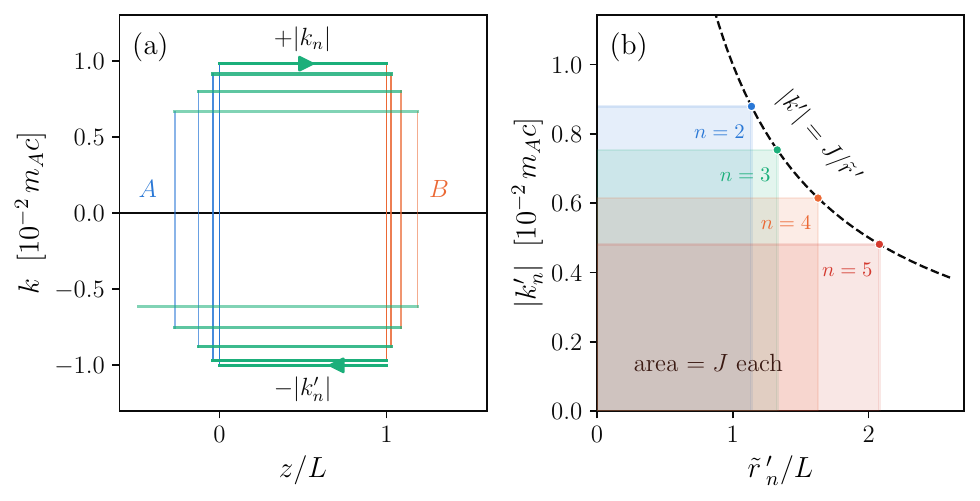}
\caption{\label{fig:phasespace}%
The surviving phase-space area (repulsive branch, $\mB/\mA=1.5$,
$|k_0'|=0.01\,\mA c$, blocks initially at rest, exact event-driven
dynamics). (a)~The mediator's orbit in the $(z,p)$ plane: right-moving
legs at $p=+|k_n|$, left-moving legs at $p=-|k_n'|$, reflected at the
recoiling walls ($A$ blue, $B$ orange). As the enclosure expands, the
orbit spirals outward in $z$ and inward in $p$. (b)~The geometric-mean
rectangles $\tilde r_n'\times|k_n'|$ of Eq.~(\ref{eq:exactJ}): every
corner lies on the hyperbola $|k'|=J/\tilde r\,'$ (dashed) and every
rectangle has the same area $J$ --- half the loop invariant of
Eq.~(\ref{eq:adiabatic}), surviving exactly. It is this geometric-mean
pairing that has constant area: the rectangles built on single legs,
visible in panel~(a), do not.}
\end{figure}

The null flight legs dilate by Doppler factors of the same form:
Eq.~(\ref{eq:doppler}) [or Eq.~(\ref{eq:ynxn})] gives
$(z_{n+1}'-z_{n+1})/(z_n'-z_n)=e^{-2y_n^{(A)}+2y_n^{(B)}}$, so the
geometric-mean widths (\ref{eq:gmdef}) grow per collision pair by
\begin{equation}
\frac{\tilde r_n'}{\tilde r_{n-1}'}
=e^{-\big(y_{n-1}^{(A)}+y_n^{(A)}\big)+\big(y_{n-1}^{(B)}+y_n^{(B)}\big)},
\label{eq:gmratio}
\end{equation}
exactly the inverse of the net redshift $|k_n'|/|k_{n-1}'|$
accumulated at $A_n$ and $B_n$ through Eq.~(\ref{eq:reflexact}). Hence
$|k_n'|\tilde r_n'=|k_{n-1}'|\tilde r_{n-1}'$ for every $n$, and
identically $|k_n|\tilde r_n=|k_{n-1}|\tilde r_{n-1}$; the two constants
coincide because, at equal $n$,
$\tilde r_n'/\tilde r_n=|k_n|/|k_n'|=e^{\,y_{n-1}^{(B)}+y_n^{(B)}}$. A
single collision then fixes the constant to the $J$ of
Eq.~(\ref{eq:exactJ}). The geometric-mean sampling is thus precisely the
bookkeeping that pairs each reflection factor $e^{\,y_{n-1}+y_n}$ with
its length-dilation partner: the phase-space rectangle of height $|k|$
and width $\tilde r$ --- half the loop (\ref{eq:adiabatic}) --- deforms
from tall-and-narrow to short-and-wide at
exactly constant area (Fig.~\ref{fig:phasespace}) --- the adiabatic
invariant of Eq.~(\ref{eq:adiabatic}), surviving a deformation that is
neither slow nor recoil-free.

The failure of the alternatives is itself exact. With the relative
rapidity of Sec.~\ref{sec:example},
\begin{equation}
\Delta y_n\equiv y_n^{(B)}-y_n^{(A)}=\ln(\mathcal{C}_n/\mathcal{C}_{n-1}),
\end{equation}
sampling with either single adjacent leg gives
\begin{equation}
|k_n'|\,(z_n'-z_n)=J\,e^{-\Delta y_n},
\qquad
|k_n'|\,(z_{n+1}'-z_{n+1})=J\,e^{+\Delta y_n},
\label{eq:sampling}
\end{equation}
while the arithmetic mean of the two gives $J\cosh\Delta y_n$ [by
Eqs.~(\ref{eq:knpabs}), (\ref{eq:ynxn}) and (\ref{eq:gmvalues})]. Since
$\Delta y_n$ climbs monotonically to $\theta$, these drift to
$Je^{\mp\theta}$ and $J\cosh\theta$; all sampling conventions coincide
only in the adiabatic regime $\theta\to0$. In the $(z,p)$ plane of
Fig.~\ref{fig:phasespace}(a) these products are the areas of rectangles
built on a \emph{single} flight leg --- height $|k_n'|$, width one
adjacent leg --- and they drift; only the geometric-mean rectangle of
panel (b) keeps a constant area.
The loop integral along the true trajectory, $|k_n|(z_n'-z_n)+|k_n'|(z_n'-z_{n+1})$ --- each leg weighted by its own momentum --- is likewise not conserved: it tends to $2J\cosh(2\ln\mathcal{X})$ with $\mathcal{X}=\Lambda_1e^{\theta/2}/[(\mA+\mB e^{\theta})c]$, which in the initial rest frame exceeds $2J$ by $O(\theta^{2})$.

The exactness is special to the massless mediator. A Newtonian mediator
reflects \emph{additively}, $u\to-u+2V+O(m/M)$, not multiplicatively in
an exponentiated variable, and its flight legs are not null; no pairing
of the kind above exists, and the area is conserved only adiabatically,
with the $O(m/M)$ collision-by-collision residual quantified in the
footnote of Sec.~\ref{sec:potential}. The multiplicative Doppler algebra
of null legs is what upgrades the adiabatic area to an exact invariant.

\section{The collision map: matrix form and conjugacy}
\label{app:map}

The recursion (\ref{eq:mobius}) is the projective action of a $2\times2$
matrix on the ratio $p_n^-=g_n/f_n$: written out,
$p_{n+1}^-=(ap_n^-+b)/(cp_n^-+d)$ with
\begin{equation}
M=\begin{pmatrix} a & b\\ c & d\end{pmatrix}
 =\begin{pmatrix}
   s/c^{2}-(\mB c)^{2} & -\Lambda_2(\mA c)^{2}/\sqrt2\\[2pt]
   \sqrt2\,\Lambda_1 & -(\mA c)^{2}
  \end{pmatrix},
\qquad
\begin{aligned}
\mathrm{tr}\,M&=2(\mA c)(\mB c)\,\alpha,\\
\det M&=(\mA c)^{2}(\mB c)^{2}.
\end{aligned}
\label{eq:Mmatrix}
\end{equation}
The matrix acts on $p_n^-$ through the linear system behind it: with
$p_n^-=g_n/f_n$ as in Sec.~\ref{sec:solution}, one collision pair is, up
to a common rescaling of $(g_n,f_n)$,
\begin{equation}
\begin{pmatrix} g_{n+1}\\ f_{n+1}\end{pmatrix}
=M\begin{pmatrix} g_n\\ f_n\end{pmatrix},
\qquad
p_n^-=\frac{\big[M^{\,n}(p_0^-,1)^{\mathsf T}\big]_1}
           {\big[M^{\,n}(p_0^-,1)^{\mathsf T}\big]_2},
\label{eq:Mlift}
\end{equation}
so iterating the collision map is taking powers of $M$, and rescaling
$M$ changes nothing in the ratio. Only the ratio of the eigenvalues,
\begin{equation}
\lambda_\pm=(\mA c)(\mB c)\,e^{\pm\theta},
\qquad
\frac{\lambda_+}{\lambda_-}=e^{2\theta},
\label{eq:Meig}
\end{equation}
is therefore physical, and it is fixed by $\mathrm{tr}\,M$ and $\det M$
alone, through $\mathrm{tr}\,M/\sqrt{\det M}=2\alpha$ --- the
conjugacy invariant of a M\"obius map~\cite{Beardon1991}. For
$|\alpha|>1$ the map is
\emph{hyperbolic}, conjugate to multiplication by $e^{2\theta}$ --- the
scattering regime of Secs.~\ref{sec:solution}--\ref{sec:potential}, and
the unphysical branch $\alpha\le-1$ of Sec.~\ref{sec:attractive}; for
$|\alpha|=1$ it is parabolic; for $|\alpha|<1$ it is \emph{elliptic},
conjugate to multiplication by $e^{2i\vartheta}$ --- the bound window of
Sec.~\ref{sec:attractive}, the phase $\chi_n=n\vartheta+\tilde\varphi$ of
the continued amplitudes advancing by $\vartheta$ per collision pair. In
either case the fixed points are the terminal states
[cf.~Eq.~(\ref{eq:terminal})]: real, and approached exponentially, in
the scattering regime; complex in the bound window, where no terminal
state exists and the orbit circulates forever.


\bibliography{references}

@article{Ee2012,
    author = {Ee, June-Haak and Lee, Jungil},
    title = {A unique pure mechanical system revealing dipole repulsion},
    journal = {Am. J. Phys.},
    volume = {80},
    number = {12},
    pages = {1078--1084},
    year = {2012},
    doi = {10.1119/1.4756036},
}

@article{Ee2015,
    author = {Ee, June-Haak and Lee, Jungil},
    title = {Magic mass ratios of complete energy-momentum transfer in one-dimensional elastic three-body collisions},
    journal = {Am. J. Phys.},
    volume = {83},
    number = {2},
    pages = {110--120},
    year = {2015},
    doi = {10.1119/1.4897162},
}

@article{Sinai,
    author = {Sinai, Ya. G.},
    title = {Dynamics of a heavy particle surrounded by a finite number of light particles},
    journal = {Theor. Math. Phys.},
    volume = {121},
    number = {1},
    pages = {1351--1357},
    year = {1999},
    doi = {10.1007/BF02557232},
}

@article{Galperin2003,
    author = {Galperin, G.},
    title = {Playing pool with $\pi$ (the number $\pi$ from a billiard point of view)},
    journal = {Regul. Chaotic Dyn.},
    volume = {8},
    pages = {375--394},
    year = {2003},
    doi = {10.1070/RD2003v008n04ABEH000252},
}

@article{Redner2004,
    author = {Redner, S.},
    title = {A billiard-theoretic approach to elementary one-dimensional elastic collisions},
    journal = {Am. J. Phys.},
    volume = {72},
    number = {12},
    pages = {1492--1498},
    year = {2004},
    doi = {10.1119/1.1738428},
}

@misc{3b1b_1,
    author = {Sanderson, Grant},
    title = {The most unexpected answer to a counting puzzle},
    year = {2019},
    note = {3Blue1Brown video lesson, \url{https://www.3blue1brown.com/lessons/clacks}},
}

@misc{3b1b_2,
    author = {Sanderson, Grant},
    title = {Why do colliding blocks compute $\pi$?},
    year = {2019},
    note = {3Blue1Brown video lesson, \url{https://www.3blue1brown.com/lessons/clacks-solution}},
}

@article{Brown:2019jvs,
    author = {Brown, Adam R.},
    title = {{Playing Pool with $|\psi\rangle$: from Bouncing Billiards to Quantum Search}},
    eprint = {1912.02207},
    archivePrefix = {arXiv},
    primaryClass = {quant-ph},
    doi = {10.22331/q-2020-11-02-357},
    journal = {Quantum},
    volume = {4},
    pages = {357},
    year = {2020},
}

@article{Aretxabaleta2020,
    author = {Aretxabaleta, Xabier M. and Gonchenko, Marina and Harshman, Nathan L. and Jackson, Steven Glenn and Olshanii, Maxim and Astrakharchik, Grigory E.},
    title = {The Dynamics of Digits: Calculating Pi with {G}alperin's Billiards},
    journal = {Mathematics},
    volume = {8},
    number = {4},
    pages = {509},
    year = {2020},
    doi = {10.3390/math8040509},
}

@book{LandauMechanics,
    title = {Mechanics},
    author = {Landau, Lev Davidovich and Lifshitz, Evgeni Mikhailovich},
    volume = {1},
    edition = {3},
    year = {1976},
    publisher = {Butterworth-Heinemann},
    series = {Course of Theoretical Physics},
}

@misc{Lancaster:2015lfa,
    author = {Lancaster, Jarrett L. and McGuire, Colin and Titus, Aaron P.},
    title = {{Emergence of long-range force laws from classical particle exchange}},
    eprint = {1509.02885},
    archivePrefix = {arXiv},
    primaryClass = {physics.class-ph},
    year = {2015},
}

@article{Artigue:2018jwu,
    author = {Artigue, Alfonso},
    title = {{Billiards and toy gravitons}},
    eprint = {1808.10547},
    archivePrefix = {arXiv},
    primaryClass = {math-ph},
    doi = {10.1007/s10955-019-02252-0},
    journal = {J. Stat. Phys.},
    volume = {175},
    pages = {213--232},
    year = {2019},
}

@article{ArtigueRel2024,
    author = {Artigue, Alfonso},
    title = {{Relativistic one-dimensional billiards}},
    eprint = {2504.00166},
    archivePrefix = {arXiv},
    primaryClass = {math.DS},
    doi = {10.1007/s10955-024-03364-y},
    journal = {J. Stat. Phys.},
    volume = {191},
    pages = {156},
    year = {2024},
}

@misc{inprep,
    author = {Ee, June-Haak and Kim, U-Rae and Lee, Jungil},
    year = {2026},
    note = {in preparation},
}

@book{Beardon1991,
    author = {Beardon, Alan F.},
    title = {Iteration of Rational Functions: Complex Analytic Dynamical Systems},
    series = {Graduate Texts in Mathematics},
    volume = {132},
    publisher = {Springer},
    address = {New York},
    year = {1991},
}

@article{Fermi1949,
    author = {Fermi, Enrico},
    title = {On the Origin of the Cosmic Radiation},
    journal = {Phys. Rev.},
    volume = {75},
    pages = {1169--1174},
    year = {1949},
    doi = {10.1103/PhysRev.75.1169},
}

@inproceedings{Ulam1961,
    author = {Ulam, S. M.},
    title = {On some statistical properties of dynamical systems},
    booktitle = {Proc. Fourth Berkeley Symp. on Math. Statist. and Prob.},
    volume = {3},
    pages = {315--320},
    publisher = {Univ. of California Press},
    address = {Berkeley},
    year = {1961},
}

@article{Schwinger1962,
    author = {Schwinger, Julian},
    title = {Gauge Invariance and Mass. {II}},
    journal = {Phys. Rev.},
    volume = {128},
    pages = {2425--2429},
    year = {1962},
    doi = {10.1103/PhysRev.128.2425},
}

@article{BlochNordsieck1937,
    author = {Bloch, F. and Nordsieck, A.},
    title = {Note on the Radiation Field of the Electron},
    journal = {Phys. Rev.},
    volume = {52},
    pages = {54--59},
    year = {1937},
    doi = {10.1103/PhysRev.52.54},
}

@incollection{GruberLesne2006,
    author = {Gruber, Christian and Lesne, Annick},
    title = {Adiabatic Piston},
    booktitle = {Encyclopedia of Mathematical Physics},
    editor = {Fran{\c{c}}oise, Jean-Pierre and Naber, Gregory L. and Tsun, Tsou Sheung},
    publisher = {Elsevier},
    address = {Oxford},
    pages = {160--174},
    year = {2006},
}

@article{Ehrenfest1916,
    author = {Ehrenfest, P.},
    title = {Adiabatische {I}nvarianten und {Q}uantentheorie},
    journal = {Ann. Phys.},
    volume = {356},
    pages = {327--352},
    year = {1916},
    doi = {10.1002/andp.19163561905},
}

@article{Coleman1976,
    author = {Coleman, Sidney},
    title = {More about the massive {S}chwinger model},
    journal = {Ann. Phys.},
    volume = {101},
    pages = {239--267},
    year = {1976},
    doi = {10.1016/0003-4916(76)90280-3},
}

@misc{SkinnerNeishtadt2024,
    author = {Skinner, Joshua and Neishtadt, Anatoly},
    title = {{Unusual properties of adiabatic invariance in a billiard model
              related to the adiabatic piston problem}},
    eprint = {2409.07458},
    archivePrefix = {arXiv},
    primaryClass = {math.DS},
    year = {2024},
}

@article{NielsenOlsen1966,
    author = {Nielsen, A. and Olsen, J.},
    title = {Formal analogy between {C}ompton scattering and {D}oppler effect},
    journal = {Am. J. Phys.},
    volume = {34},
    number = {7},
    pages = {621--622},
    year = {1966},
    doi = {10.1119/1.1973153},
}

@article{Kidd1985,
    author = {Kidd, Richard and Ardini, James and Anton, Anatol},
    title = {{C}ompton effect as a double {D}oppler shift},
    journal = {Am. J. Phys.},
    volume = {53},
    number = {7},
    pages = {641--644},
    year = {1985},
    doi = {10.1119/1.14274},
}

@article{CaiZhang2023,
    author = {Cai, Yin and Zhang, Fu-Lin},
    title = {Hear $\pi$ from quantum {G}alperin billiards},
    journal = {Can. J. Phys.},
    volume = {101},
    number = {9},
    pages = {491--495},
    year = {2023},
    doi = {10.1139/cjp-2022-0262},
}

@book{BLP,
    author = {Berestetskii, Vladimir B. and Lifshitz, Evgeni M. and Pitaevskii, Lev P.},
    title = {Quantum Electrodynamics},
    volume = {4},
    edition = {2},
    year = {1982},
    publisher = {Butterworth-Heinemann},
    series = {Course of Theoretical Physics},
    note = {\S 86},
}

@article{Moore1970,
    author = {Moore, Gerald T.},
    title = {Quantum theory of the electromagnetic field in a
             variable-length one-dimensional cavity},
    journal = {J. Math. Phys.},
    volume = {11},
    number = {9},
    pages = {2679--2691},
    year = {1970},
    doi = {10.1063/1.1665432},
}

@article{Passon2018,
    author = {Passon, Oliver and Z{\"u}gge, Thomas and Grebe-Ellis, Johannes},
    title = {{Pitfalls in the teaching of elementary particle physics}},
    journal = {Phys. Educ.},
    volume = {54},
    number = {1},
    pages = {015014},
    year = {2019},
    doi = {10.1088/1361-6552/aadbc7},
    eprint = {1811.06230},
    archivePrefix = {arXiv},
    primaryClass = {physics.ed-ph},
}

\end{document}